# Deep learning based CT-to-CBCT deformable image registration for autosegmentation in head and neck adaptive radiation therapy

Xiao Liang, Howard Morgan, Dan Nguyen, Steve Jiang[*]

Medical Artificial Intelligence and Automation Laboratory and Department of Radiation Oncology, University of Texas Southwestern Medical Center, Dallas, TX, USA
[*]E-mail: Steve.Jiang@UTSouthwestern.edu
[*]Address: 2280 Inwood Road, Dallas, Texas, 75390-9303
[*]Phone: (+1)214-645-8510

**Abstract**

The purpose of this study is to develop a deep learning based method that can automatically generate segmentations on cone-beam CT (CBCT) for head and neck online adaptive radiation therapy (ART), where expert-drawn contours in planning CT (pCT) can serve as prior knowledge. Due to lots of artifacts and truncations on CBCT, we propose to utilize a learning based deformable image registration method and contour propagation to get updated contours on CBCT. Our method takes CBCT and pCT as inputs, and output deformation vector field and synthetic CT (sCT) at the same time by jointly training a CycleGAN model and 5-cascaded Voxelmorph model together. The CycleGAN serves to generate sCT from CBCT, while the 5-cascaded Voxelmorph serves to warp pCT to sCT's anatomy. The segmentation results were compared to Elastix, Voxelmorph and 5-cascaded Voxelmorph on 18 structures including left brachial plexus, right brachial plexus, brainstem, oral cavity, middle pharyngeal constrictor, superior pharyngeal constrictor, inferior pharyngeal constrictor, esophagus, nodal gross tumor volume, larynx, mandible, left masseter, right masseter, left parotid gland, right parotid gland, left submandibular gland, right submandibular gland, and spinal cord. Results show that our proposed method can achieve average Dice similarity coefficients of 0.83±0.09 and average 95% Hausdorff distance of 2.01±1.81 mm. As compared to other methods, our method has shown better accuracy to Voxelmorph and 5-cascaded Voxelmorph, and comparable accuracy to Elastix but much higher efficiency. The proposed method can rapidly and simultaneously generate sCT with correct CT numbers and propagate contours from pCT to CBCT for online ART re-planning.

## 1. Introduction

Studies have shown that adaptive radiation therapy (ART) can improve the dosimetric quality of radiation therapy plans by altering the treatment plans based on patient anatomical changes [1]. However, the time-consuming parts in ART including segmentation and re-planning make online ART difficult to implement in clinics. Recently, several commercially available online ART systems have been developed: Ethos ™ (Varian Inc., Palo Alto, USA), MRIdian ™ (ViewRay Inc., Cleveland, OH, USA), and Unity ™ (Elekta AB Inc., Stockholm, Sweden). Ethos [2] is a CBCT-based online ART platform that works with Halcyon Linac, while MRIdian [3] and Unity [4] are MRI-based online ART platforms that work with MRI Linacs.

Even though MRI images have much better soft tissue contrast than CBCT images, CBCT images will still be often used in ART as MRI is unsuitable to patients with metal implants because of the magnetic field and to clinics with value-based health care because of its expensive price. Among tumor sites that often have inter-fractional anatomical changes during RT, head-and-neck (H&N) cancer is one of the sites that can benefits from CBCT-based online ART. A clinical study of ART benefits on H&N patients showed that dose to parotid gland were significantly reduced for all 30 patients [5]. Another study showed that target coverage for patients who underwent adaptation of their treatment plans improved by up to 10.7% of median dose [6]. Thus utilizing CBCT in ART workflow can avoid the risk of tumor underdose and organs-at-risk (OAR) overdose.

To use CBCT in ART workflow, corrections need to be done on CBCT. Compared to CT, CBCT has lots of artifacts and inaccurate Hounsfield Units (HU) values. In order to get accurate dose calculation on CBCT, HU values need to be corrected and artifacts need to be removed on CBCT. Our previous work used CycleGAN which is a deep learning based method to convert CBCT to synthetic CT (sCT) images that have CT's HU values and reduced artifacts, and the dose distributions calculated on sCT showed a higher gamma index pass rate than those on CBCT [7]. Deep learning introduces an easier, faster and more accurate way to generate sCT from CBCT for dose calculation in ART compared with deformable image registration method (DIR), because 1. It doesn't require paired data for training; 2. It enables rapid deployment of a trained mode; 3. It preserves CBCT's anatomy.

Besides accurate dose calculation, another problem of using CBCT in online ART is how to achieve accurate autosegmentation. Due to lots of artifacts and axial truncation on CBCT of H&N site, directly using deep learning method to contour OARs and target on CBCT images is very challenging. One study used CycleGAN to convert CBCT to synthetic MRI images, and then combined the CBCT and synthetic MRI together to enhance the training of deep learning based multi-organs autosegmentation model [8]. Most studies of getting autosegmentation from CBCT for online ART and the state-of-the-art methods still focus on DIR based methods to get deformation vector field (DVF) from warping planning CT (pCT) to CBCT's anatomy, and then applying DVF to the contours on pCT to get the updated contours on CBCT [9]. However, DIR could generate inaccurate segmentations because of more pronounced anatomical changes and low soft tissue contrast [10].

It is reasonable to use contour propagation for autosegmentation in online ART in that DIR methods leverage prior knowledge from contours on pCT. DIR between daily/weekly CBCT and planning CT is often used in H&N ART workflow to get the most up-to-date anatomy. So far, the processes of CBCT-to-sCT conversion and pCT-to-CBCT DIR are always done separately. Therefore, we propose a method combining a CycleGAN model and a deep-learning-based DIR model together and jointly training them. The CycleGAN model is used to convert CBCT to sCT images, and the generated sCT is used by the DIR model to do same modality image registration (sCT-to-CT) rather than different modality image registration (CBCT-to-CT), where DIR is considered more accurate within the same image modality than cross different modalities [11]. The DIR model generates DVFs and deformed planning CT (dpCT) by deforming pCT to sCT's anatomy, and the generated dpCT is used to guide CycleGAN to generate more accurate sCT from CBCT during the CycleGAN training. A better quality sCT then leads to more accurate image registration. In this way, the two deep learning models can interactively improve each other than training them alone. This method can also generate sCT from CBCT and updated contours from contour propagation at the same time for ART.

## 2. Data

We retrospectively collected data from 124 patients with head and neck squamous cell carcinoma treated with external beam radiotherapy with radiation dose around 70Gy. Each patient includes a pCT volume acquired before the treatment course, OARs and target segmentations delineated by physicians on the pCT, and a CBCT volume acquired around fraction 20 during treatment course. The pCT volumes were acquired by a Philips CT scanner with $1.17 \times 1.17 \times 3.00$ mm$^3$ voxel spacing. The CBCT volumes were acquired by Varian On-Board Imagers with $0.51 \times 0.51 \times 1.99$ mm$^3$ voxel spacing and $512 \times 512 \times 93$ dimensions. The pCT is rigid registered to CBCT through Velocity (Varian Inc., Palo Alto, USA). Therefore the rigid-registered pCT has the same voxel spacing and dimensions with CBCT. After rigid registration, the dimensions of pCT and CBCT volumes were both downsampled to $256 \times 256 \times 93$ from $512 \times 512 \times 93$, and then cropped to $224 \times 224 \times 80$. We divided the dataset to 100/4/20 for training/validation/testing respectively. Since our proposed model is an unsupervised learning method, contours on pCT and ground truth contours on CBCT are not needed for training. However, ground truth contours were needed to evaluate the accuracy of the proposed DIR network. We selected 18 structures that either are critical OARs or have large anatomical changes during radiotherapy courses. They are left brachial plexus, right brachial plexus, brainstem, oral cavity, middle pharyngeal constrictor, superior pharyngeal constrictor, inferior pharyngeal constrictor, esophagus, nodal gross tumor volume (nGTV), larynx, mandible, left masseter, right masseter, left parotid gland, right parotid gland, left submandibular gland, right submandibular gland, spinal cord. The contours of the these 18 structures were first propagated from pCT to CBCT using rigid and deformable image registration in Velocity, and then modified and approved by an radiation oncology expert to obtain the ground truth for validation and testing.

## 3. Methods

### 3.1. Overview of CycleGAN

Our previous study used CycleGAN architecture to convert CBCT to sCT images that have less artifacts and accurate HU values [7]. The architecture we used in this study is shown in Figure 1. It has two generators with U-Net architecture and two discriminators with patchGAN architecture. $G_A$ generates sCT from CBCT, and $G_B$ generates sCBCT from CT. $D_A$ is used to distinguish between sCT and CT, while $D_B$ is used to distinguish between sCBCT and CBCT. There are two cycles in the CycleGAN: 1. $G_A$ takes CBCT as input and outputs sCT, then $G_B$ takes sCT as input and outputs cycle CBCT (cCBCT); 2. $G_B$ takes CT as input and outputs sCBCT, then $G_A$ takes sCBCT as input and outputs cycle CT (cCT). Even though $G_A$ is used to output sCT from CBCT, it still can generate identical CT (iCT) if the input to it is CT, and vice versa. In summary, the loss function for generators is

$$\mathcal{L}_G = \mathcal{L}_{GAN-G} + \alpha \times \mathcal{L}_{Cycle} + \beta \times \mathcal{L}_{Identity}, \quad (1)$$

$$\mathcal{L}_{GAN-G} = \frac{1}{m}\sum_{i=1}^{m}((1 - D_A(sCT_i))^2 + (1 - D_B(sCBCT_i))^2), \quad (2)$$

$$\mathcal{L}_{Cycle} = \frac{1}{m}\sum_{i=1}^{m}(|cCBCT_i - CBCT_i| + |cCT_i - CT_i|), \quad (3)$$

$$\mathcal{L}_{Identity} = \frac{1}{m}\sum_{i=1}^{m}(|iCT_i - CT_i| + |iCBCT_i - CBCT_i|), \quad (4)$$

The loss function for discriminators is

$$\mathcal{L}_D = \mathcal{L}_{GAN-D}, \quad (5)$$

$$\mathcal{L}_{GAN-D} = \frac{1}{m}\sum_{i=1}^{m}\left(\frac{(1-D_A(CT_i))^2+(D_A(sCT_i))^2}{2} + \frac{(1-D_B(CBCT_i))^2+(D_B(sCBCT_i))^2}{2}\right). \quad (6)$$

For more details, the reader is kindly referred to Liang *et al* [7].

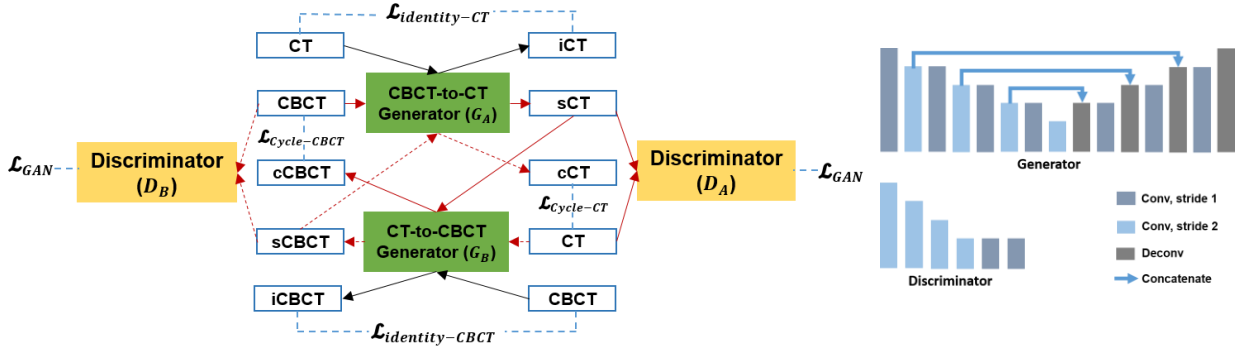

**Figure 1.** CycleGAN model architecture. The left figure is the architecture of CycleGAN, and the right figure is the neural networks of generators and discriminators used in the CycleGAN. Red dotted arrows illustrate the flow of one cycle, and red solid arrows illustrate the flow of another cycle. Blue dashed lines connect the values used in loss function.

### 3.2. Overview of Voxelmorph

Recently, learning based DIR methods have gained lots of attention for its fast deployment compared to classical DIR techniques. One of the state-of-the-art networks is Voxelmorph [12]. This model assumes that DVF can be defined by the following ordinary differential equation (ODE):

$$\frac{\partial \phi^{(t)}}{\partial t} = z(\phi^{(t)}), \quad (7)$$

where $t$ is time, $z$ is velocity field, and $\phi$ is DVF. $\phi^{(0)} = Id$ is the identity transformation. $\phi^{(1)}$ is the final registration field obtained by integrating the stationary velocity field $z$ over $t = [0,1]$. In this way, deformations are diffeomorphic, differentiable, and invertible, which can preserve topology. Given this assumption, Voxelmorph takes moving images pCT ($I_m$) and fix images sCT ($I_f$) as input, and outputs voxel-wise mean ($\mu_{z|I_m,I_f}$) and variance ($\Sigma_{z|I_m,I_f}$) of velocity field with U-Net architecture, shown in Figure 2(a). Then velocity field $z$ is sampled from predicted $\mu_{z|I_m,I_f}$ and $\Sigma_{z|I_m,I_f}$ with the following equation:

$$z = \mu_{z|I_m,I_f} + \sqrt{\Sigma_{z|I_m,I_f}}\, r, \tag{8}$$

where r is a sample from the standard normal distribution: $r \sim \mathcal{N}(0, I)$. Given velocity field $z$, DVF ($\phi_z$) can be calculated with scaling and squaring operations. Finally, a spatial transform layer is integrated to warp pCT to sCT's anatomy using the predicted DVF to get dpCT ($I'_m$). New contours can also be calculated with pCT contours and predicted DVF through the spatial transform layer. The loss function of the Voxelmorph architecture is

$$\mathcal{L}_V = \mathcal{L}_R + \mathcal{L}_{Image\_Similarity}, \tag{9}$$

$$\mathcal{L}_R = \frac{1}{2}(tr\left(\lambda D\Sigma_{z|I_m,I_f} - log\Sigma_{z|I_m,I_f}\right) + \mu^T_{z|I_m,I_f} \Lambda_z \mu_{z|I_m,I_f}), \tag{10}$$

$$\mathcal{L}_{Image\_Similarity} = \frac{1}{2\sigma^2 m}\sum_{i=1}^{m}\left\|I'_m - I_f\right\|^2, \tag{11}$$

where $\mathcal{L}_R$ is derived from Kullback–Leibler divergence of posterior probability $p(z|I_m; I_f)$ and approximate posterior probability $\mathcal{N}(z; \mu_{z|I_m,I_f}, \Sigma_{z|I_m,I_f})$. $\mathcal{L}_R$ encourages the posterior to be close to the prior $p(z)$, and $\mathcal{L}_{Image\_Similarity}$ encourages the warped images to be similar to fix images. For more details, the reader is kindly referred to Dalca *et al* [12].

### 3.3. 5-cascaded Voxelmorph

Recursive cascaded networks for DIR have been shown to significantly outperform state-of-the-art learning based DIR methods [13]. Therefore, we pick 5-cascaded Voxelmorph network in this study to gain better DIR performance, and its architecture is shown in Figure 2(b). The input of 5-cascaded Voxelmorph is also pCT ($I_m$) and sCT ($I_f$), and we cascade Voxelmorph by successively performing DIR between warped images ($I_m^{(n)}$) and fix images ($I_f$). Each cascade predicts a new DVF between fix images and previously predicted warped images. Thus, the final DVF is a composition of all predicted DVFs:

$$\phi_z = \phi_{z_5} \circ \phi_{z_4} \circ \cdots \circ \phi_{z_1}. \tag{12}$$

The final warped images is constructed by

$$I_m^{(5)} = \phi_z \circ I_m. \tag{13}$$

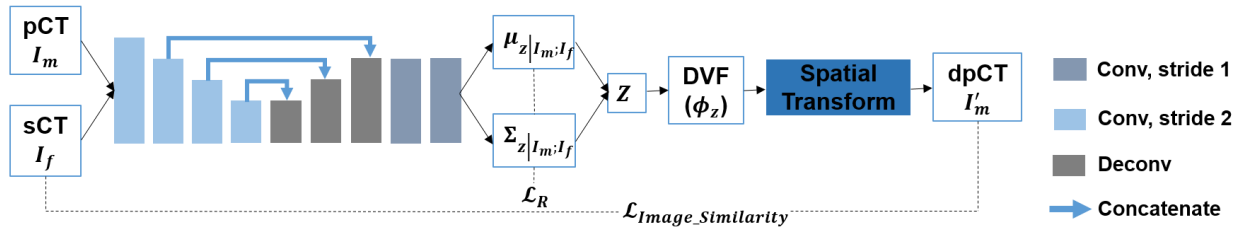

(a) Voxelmorph

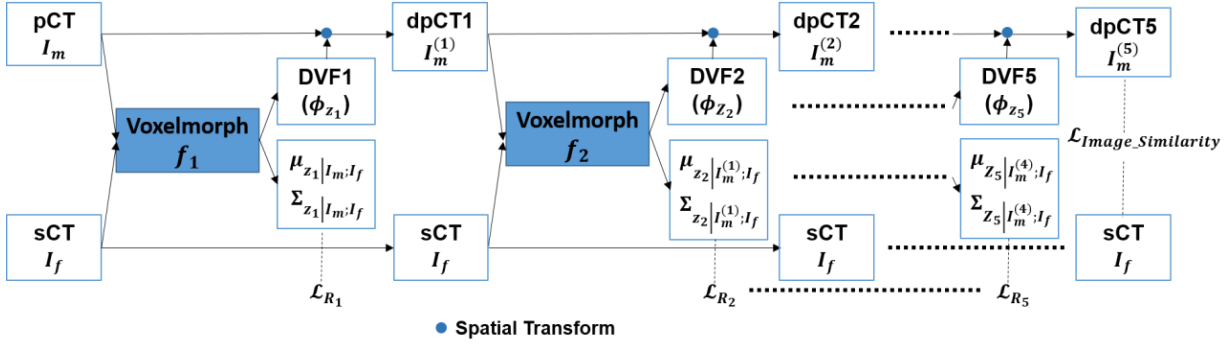

**(b) 5-cascaded Voxelmorph**

**Figure 2.** The architecture of Voxelmorph and 5-cascaded Voxelmorph. The unbold black dotted lines illustrate the loss function.

### 3.4. Joint model

The performance of DIR highly depends on the image quality of fix and moving images. In our case, the image quality of fix images, which have lots of artifacts and different HU range from moving images, is the main factor causing DIR error. Thus we use sCT generated by CycleGAN, which has less artifacts and similar range of CT HU values, to take place of CBCT as fix images. We propose a combined model by jointly training CycleGAN and 5-cascaded Voxelmorph together, shown in Figure 3. With independently pre-trained CycleGAN and 5-cascaded Voxelmorph, we firstly update the parameters of CycleGAN generators using loss function of

$$\mathcal{L}_G = \mathcal{L}_{GAN-G} + \alpha \times \mathcal{L}_{Cycle} + \beta \times \mathcal{L}_{Identity} + \gamma \times \mathcal{L}_{Similarity}, \quad (14)$$

$$\mathcal{L}_{Similarity} = \frac{1}{k}\Sigma_k |I_f - I_m^{(5)}|. \quad (15)$$

Different from training CycleGAN alone, dpCT adds another supervision to guide CycleGAN generating more realistic sCT images by adding $\mathcal{L}_{Similarity}$ in the joint model. Then we update the parameters of CycleGAN discriminators with updated synthetic and real images. Finally we update the parameters of 5-cascaded Voxelmorph with loss function of

$$\mathcal{L}_V = \mathcal{L}_{R_1} + \mathcal{L}_{R_2} + \mathcal{L}_{R_3} + \mathcal{L}_{R_4} + \mathcal{L}_{R_5} + \mathcal{L}_{Image\_Similarity}. \quad (16)$$

The more realistic sCT images CycleGAN can generate, the more accurate registration DIR can perform. By jointly training CycleGAN and Voxelmorph together, these two networks can help to improve each other than training them alone separately.

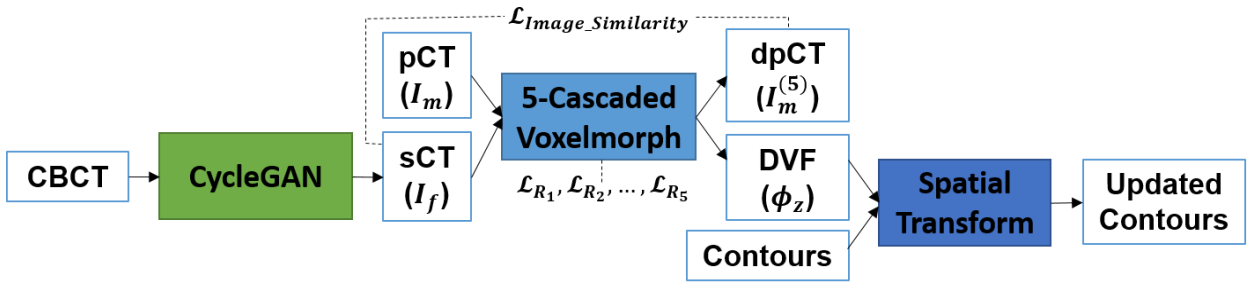

---
Algorithm: Jointly training CycleGAN and 5-cascaded Voxelmorph
---
1: pretrain CycleGAN
2: pretrain 5_cascaded Voxelmorph
3: for i =1, . . . . . do
4:     train CycleGAN generators with $\mathcal{L}_G = \mathcal{L}_{GAN-G} + \alpha \times \mathcal{L}_{Cycle} + \beta \times \mathcal{L}_{Identity} + \gamma \times \mathcal{L}_{Similarity}$
5:     train CycleGAN discriminators with $\mathcal{L}_D = \mathcal{L}_{GAN-D}$
6:     train 5_cascaded Voxelmorph with $\mathcal{L}_V = \mathcal{L}_{R_1} + \mathcal{L}_{R_2} + \mathcal{L}_{R_3} + \mathcal{L}_{R_4} + \mathcal{L}_{R_5} + \mathcal{L}_{Image\_Similarity}$
---

**Figure 3.** The architecture of the joint model and its training scheme. Black dashed lines illustrate loss function.

### 3.5. Training details

The CycleGAN, Voxelmorph, 5-cascaded Voxelmorph, and joint model were all trained with volume size of 224 × 224 × 80 on a NVIDIA Tesla V100 GPU with memory of 32GB. The maximum cascades we can have for the volume size of 224 × 224 × 80 and batch size of 1 without exceeding GPU memory capacity is 5. Adam optimization with learning rate of 0.0002 was used for all the models training. Hyperparameters α, β, $\sigma^2$, and $\lambda$ were set to 10, 5, 0.02, and 30. The learning rate and above hyperparameters were set to the same values as in previous studies. We found that joint model performs the best with $\gamma=10$.

### 3.6. Evaluation methods

To quantitatively evaluate segmentation accuracy, multiple metrics including dice similarity coefficient (DSC), relative absolute volume difference (RAVD), and 95% Hausdorff distantace (HD95) were calculated. DSC is intended to gauge the similarity of the prediction and ground truth by measuring volumetric overlap between them. It is defined as

$$DSC = \frac{2|X \cap Y|}{|X|+|Y|}, \tag{17}$$

where X is the prediction, and Y is the ground truth.

Like DSC, RAVD also measures volumetric discrepancies between the ground truth and predicted segmentation. RAVD is defined as

$$RAVD = \frac{|X-Y|}{Y}. \tag{18}$$

HD is the maximum distance of a set to the nearest point in the other set. It can be defined as

$$HD(X,Y) = \max(d_{XY}, d_{YX}) = \max\{\max_{x \in X} \min_{y \in Y} d(x,y), \max_{y \in Y} \min_{x \in X} d(x,y)\}. \tag{19}$$

HD95 is based on 95<sup>th</sup> percentile of the distances between boundary points in X and Y. The purpose of doing so is to avoid the impact of a small subset of the outliers.

## 4. Results

We compared our proposed method, which is joint model, with Voxelmorph and 5-cascaded Voxelmorph. We also compared the joint model with a non-learning based state-of-the-art method - Elastix. Elastix is a publicly available intensity-based medical image registration toolbox, extended from ITK [14]. The quantitative evaluation results including DSC, RAVD, and HD95 between predicted contours and ground truth contours of the 18 structures for 20 test patients are shown in Table 1. Our proposed method achieved DSC of 0.61, 0.63, 0.94, 0.94, 0.75, 0.72, 0.85, 0.80, 0.81, 0.89, 0.88, 0.91, 0.92, 0.91, 0.86, 0.79, 0.78, and 0.89 for left brachial plexus, right brachial plexus, brainstem, oral cavity, middle pharyngeal constrictor, superior pharyngeal constrictor, inferior pharyngeal constrictor, esophagus, nGTV, larynx, mandible, left masseter, right masseter, left parotid gland, right parotid gland, left submandibular gland, right submandibular gland, and spinal cord respectively. Paired student t-tests were calculated for all the metrics for statistical analysis. The calculated p-values are shown in Table 2. From Table 1 and Table 2, we found that our proposed method outperforms Voxelmorph for all the structures except left brachial plexus and right brachial plexus. When compared to 5-cascaded Voxelmorph, our proposed method has better performance on brainstem, oral cavity, middle pharyngeal constrictor, mandible, left masseter, right masseter, left parotid land, right parotid gland, and right submandibular gland in at least one evaluation metric. Our proposed method is comparable to Elastix on most of 18 structures. However, it is superior to Elastix on mandible, esophagus, and superior pharyngeal constrictor, and inferior to Elastix on left and right brachial plexus. For visual evaluation, segmentations of two test patients from Elastix, Voxelmorph, 5-cascaded Voxelmorph, joint model, and ground truth are shown in Figure 4 and Figure 5 respectively, where similar phenomenon can be observed.

**Table 1.** Quantitative evaluation of segmentations by Elastix, Voxelmorph, 5-cascaded Voxelmorph, and joint model with DSC, RAVD, and HD95 matrics. The values in the table are mean±SD.

| Structure | Method | DSC | RAVD | HD95 (mm) |
|---|---|---|---|---|
| Left brachial plexus | Elastix | 0.80±0.08 | 0.10±0.07 | 1.27±0.56 |
| | Voxelmorph | 0.60±0.18 | 0.14±0.10 | 3.59±2.01 |
| | 5-cascaded Voxelmorph | 0.65±0.16 | 0.10±0.08 | 3.10±1.48 |
| | Joint model | 0.61±0.18 | 0.12±0.12 | 3.07±2.74 |
| Right brachial plexus | Elastix | 0.80±0.08 | 0.10±0.07 | 1.36±0.70 |
| | Voxelmorph | 0.60±0.20 | 0.12±0.09 | 3.76±1.83 |
| | 5-cascaded Voxelmorph | 0.67±0.16 | 0.09±0.07 | 3.00±1.42 |
| | Joint model | 0.63±0.18 | 0.14±0.11 | 2.92±2.53 |
| Brainstem | Elastix | 0.94±0.03 | 0.01±0.01 | 0.51±0.00 |
| | Voxelmorph | 0.89±0.05 | 0.08±0.06 | 1.94±0.60 |
| | 5-cascaded Voxelmorph | 0.90±0.04 | 0.03±0.05 | 0.69±0.58 |
| | Joint model | 0.94±0.06 | 0.01±0.02 | 0.66±0.55 |
| Oral cavity | Elastix | 0.95±0.03 | 0.02±0.02 | 1.62±0.54 |
| | Voxelmorph | 0.91±0.04 | 0.05±0.03 | 3.77±1.31 |
| | 5-cascaded Voxelmorph | 0.92±0.03 | 0.01±0.03 | 1.74±1.10 |
| | Joint model | 0.94±0.04 | 0.01±0.03 | 1.77±1.15 |
| Middle pharyngeal constrictor | Elastix | 0.73±0.08 | 0.19±0.10 | 2.62±0.98 |
| | Voxelmorph | 0.65±0.15 | 0.16±0.13 | 4.22±2.53 |
| | 5-cascaded Voxelmorph | 0.71±0.10 | 0.11±0.10 | 2.07±2.21 |
| | Joint model | 0.75±0.12 | 0.10±0.12 | 2.00±2.25 |
| Superior pharyngeal constrictor | Elastix | 0.68±0.12 | 0.27±0.13 | 2.66±1.31 |
| | Voxelmorph | 0.66±0.10 | 0.15±0.11 | 3.50±2.69 |
| | 5-cascaded Voxelmorph | 0.72±0.08 | 0.13±0.09 | 1.62±1.63 |
| | Joint model | 0.72±0.08 | 0.12±0.12 | 1.78±2.13 |
| Inferior pharyngeal constrictor | Elastix | 0.83±0.06 | 0.13±0.10 | 2.23±0.52 |
| | Voxelmorph | 0.72±0.13 | 0.17±0.14 | 3.98±1.94 |
| | 5-cascaded Voxelmorph | 0.83±0.10 | 0.13±0.14 | 2.10±1.33 |
| | Joint model | 0.85±0.12 | 0.12±0.13 | 2.11±1.86 |
| Esophagus | Elastix | 0.85±0.08 | 0.09±0.08 | 1.77±0.36 |
| | Voxelmorph | 0.75±0.15 | 0.19±0.16 | 3.33±2.11 |
| | 5-cascaded Voxelmorph | 0.80±0.10 | 0.11±0.08 | 1.69±0.95 |
| | Joint model | 0.80±0.10 | 0.09±0.08 | 1.67±1.67 |
| nGTV | Elastix | 0.81±0.07 | 0.13±0.13 | 2.53±1.07 |
| | Voxelmorph | 0.67±0.12 | 0.36±0.21 | 6.19±3.38 |
| | 5-cascaded Voxelmorph | 0.82±0.11 | 0.12±0.22 | 2.83±3.52 |
| | Joint model | 0.81±0.11 | 0.13±0.22 | 2.89±3.51 |
| Larynx | Elastix | 0.89±0.06 | 0.07±0.10 | 3.60±2.75 |
| | Voxelmorph | 0.82±0.11 | 0.07±0.05 | 5.53±3.30 |
| | 5-cascaded Voxelmorph | 0.88±0.08 | 0.06±0.06 | 3.70±2.99 |

| Structure | Method | DSC | RAVD | HD95 |
|---|---|---|---|---|
| Mandible | Joint model | 0.89±0.09 | 0.07±0.06 | 3.66±3.22 |
| | Elastix | 0.87±0.05 | 0.15±0.10 | 2.19±0.82 |
| | Voxelmorph | 0.85±0.06 | 0.21±0.12 | 2.48±1.05 |
| | 5-cascaded Voxelmorph | 0.87±0.04 | 0.18±0.11 | 1.97±0.92 |
| | Joint model | 0.88±0.05 | 0.14±0.10 | 1.88±0.97 |
| Left masseter | Elastix | 0.90±0.04 | 0.05±0.03 | 1.49±0.39 |
| | Voxelmorph | 0.86±0.04 | 0.06±0.05 | 2.39±0.69 |
| | 5-cascaded Voxelmorph | 0.87±0.03 | 0.06±0.05 | 1.46±0.66 |
| | Joint model | 0.91±0.03 | 0.04±0.06 | 1.36±0.52 |
| Right masseter | Elastix | 0.91±0.03 | 0.04±0.04 | 1.53±0.41 |
| | Voxelmorph | 0.87±0.04 | 0.11±0.09 | 2.36±0.75 |
| | 5-cascaded Voxelmorph | 0.89±0.03 | 0.09±0.07 | 1.19±0.34 |
| | Joint model | 0.92±0.02 | 0.08±0.07 | 1.20±0.50 |
| Left parotid gland | Elastix | 0.89±0.07 | 0.06±0.07 | 1.67±0.92 |
| | Voxelmorph | 0.81±0.07 | 0.18±0.18 | 4.18±2.37 |
| | 5-cascaded Voxelmorph | 0.88±0.07 | 0.08±0.11 | 1.03±2.37 |
| | Joint model | 0.91±0.07 | 0.07±0.10 | 1.06±2.36 |
| Right parotid gland | Elastix | 0.88±0.08 | 0.07±0.08 | 1.93±0.99 |
| | Voxelmorph | 0.77±0.09 | 0.26±0.21 | 5.01±2.21 |
| | 5-cascaded Voxelmorph | 0.86±0.09 | 0.08±0.05 | 1.92±2.26 |
| | Joint model | 0.86±0.09 | 0.07±0.05 | 1.76±2.20 |
| Left submandibular gland | Elastix | 0.81±0.11 | 0.10±0.09 | 2.13±0.76 |
| | Voxelmorph | 0.74±0.12 | 0.20±0.14 | 3.77±1.57 |
| | 5-cascaded Voxelmorph | 0.79±0.12 | 0.15±0.11 | 2.63±1.74 |
| | Joint model | 0.79±0.13 | 0.15±0.13 | 2.66±1.78 |
| Right submandibular gland | Elastix | 0.83±0.09 | 0.11±0.13 | 2.25±1.10 |
| | Voxelmorph | 0.70±0.13 | 0.20±0.19 | 4.15±1.99 |
| | 5-cascaded Voxelmorph | 0.78±0.12 | 0.14±0.11 | 2.97±1.98 |
| | Joint model | 0.78±0.13 | 0.14±0.22 | 2.82±1.93 |
| Spinal cord | Elastix | 0.91±0.04 | 0.01±0.02 | 0.74±0.99 |
| | Voxelmorph | 0.85±0.04 | 0.10±0.10 | 2.58±1.08 |
| | 5-cascaded Voxelmorph | 0.89±0.04 | 0.04±0.05 | 0.92±0.86 |
| | Joint model | 0.89±0.04 | 0.03±0.05 | 0.95±0.70 |
| Average | Elastix | 0.85±0.07 | 0.09±0.08 | 1.89±0.84 |
| | Voxelmorph | 0.76±0.10 | 0.16±0.12 | 3.71±1.86 |
| | 5-cascaded Voxelmorph | 0.82±0.08 | 0.10±0.09 | 2.04±1.57 |
| | Joint model | 0.83±0.09 | 0.09±0.10 | 2.01±1.81 |

**Table 2**. P-values of paired student t-tests between Elastix vs. joint model, Voxelmorph vs. joint model, and 5-cascaded Voxelmorph vs. joint model.

| Structure | Method | DSC | RAVD | HD95 |
|---|---|---|---|---|
| Left brachial plexus | Elastix vs. joint model | <0.01 | 0.46 | <0.01 |
| | Voxelmorph vs. joint model | 0.36 | 0.52 | 0.30 |
| | 5-cascaded Voxelmorph vs. joint model | <0.01 | 0.28 | 0.96 |
| Right brachial plexus | Elastix vs. joint model | <0.01 | 0.20 | 0.02 |
| | Voxelmorph vs. joint model | 0.07 | 0.59 | 0.11 |
| | 5-cascaded Voxelmorph vs. joint model | <0.01 | 0.09 | 0.88 |
| Brainstem | Elastix vs. joint model | 0.73 | 0.46 | 0.26 |
| | Voxelmorph vs. joint model | <0.01 | <0.01 | <0.01 |
| | 5-cascaded Voxelmorph vs. joint model | <0.01 | 0.04 | 0.81 |
| Oral cavity | Elastix vs. joint model | 0.57 | 0.48 | 0.49 |
| | Voxelmorph vs. joint model | <0.01 | <0.01 | <0.01 |
| | 5-cascaded Voxelmorph vs. joint model | <0.01 | 0.66 | 0.43 |
| Middle pharyngeal constrictor | Elastix vs. joint model | 0.31 | 0.05 | 0.20 |
| | Voxelmorph vs. joint model | <0.01 | 0.04 | <0.01 |
| | 5-cascaded Voxelmorph vs. joint model | 0.03 | 0.58 | 0.93 |
| Superior pharyngeal constrictor | Elastix vs. joint model | 0.13 | <0.01 | 0.14 |
| | Voxelmorph vs. joint model | <0.01 | 0.09 | <0.01 |
| | 5-cascaded Voxelmorph vs. joint model | 0.80 | 0.72 | 0.37 |
| Inferior pharyngeal constrictor | Elastix vs. joint model | 0.64 | 0.97 | 0.74 |
| | Voxelmorph vs. joint model | <0.01 | 0.06 | <0.01 |
| | 5-cascaded Voxelmorph vs. joint model | 0.12 | 0.91 | 0.97 |
| Esophagus | Elastix vs. joint model | 0.02 | 0.25 | 0.80 |
| | Voxelmorph vs. joint model | 0.03 | 0.06 | <0.01 |
| | 5-cascaded Voxelmorph vs. joint model | 0.59 | 0.65 | 0.95 |
| nGTV | Elastix vs. joint model | 0.96 | 0.96 | 0.73 |
| | Voxelmorph vs. joint model | <0.01 | <0.01 | <0.01 |
| | 5-cascaded Voxelmorph vs. joint model | 0.46 | 0.12 | 0.29 |
| Larynx | Elastix vs. joint model | 0.88 | 0.71 | 0.89 |
| | Voxelmorph vs. joint model | <0.01 | 0.47 | <0.01 |
| | 5-cascaded Voxelmorph vs. joint model | 0.55 | 0.60 | 0.84 |
| Mandible | Elastix vs. joint model | 0.05 | 0.43 | 0.02 |
| | Voxelmorph vs. joint model | <0.01 | <0.01 | <0.01 |
| | 5-cascaded Voxelmorph vs. joint model | 0.45 | <0.01 | 0.05 |
| Left masseter | Elastix vs. joint model | 0.30 | 0.70 | 0.44 |
| | Voxelmorph vs. joint model | <0.01 | 0.03 | <0.01 |
| | 5-cascaded Voxelmorph vs. joint model | <0.01 | 0.01 | 0.23 |
| Right masseter | Elastix vs. joint model | 0.18 | 0.23 | 0.08 |
| | Voxelmorph vs. joint model | <0.01 | 0.01 | <0.01 |
| | 5-cascaded Voxelmorph vs. joint model | <0.01 | <0.01 | 0.93 |
| Left parotid gland | Elastix vs. joint model | 0.14 | 0.88 | 0.23 |
| | Voxelmorph vs. joint model | <0.01 | <0.01 | <0.01 |
| | 5-cascaded Voxelmorph vs. joint model | <0.01 | 0.36 | 0.41 |
| Right parotid gland | Elastix vs. joint model | 0.21 | 0.76 | 0.99 |
| | Voxelmorph vs. joint model | <0.01 | <0.01 | <0.01 |
| | 5-cascaded Voxelmorph vs. joint model | <0.01 | 0.74 | <0.01 |
| Left submandibular gland | Elastix vs. joint model | 0.08 | 0.18 | 0.26 |
| | Voxelmorph vs. joint model | <0.01 | 0.23 | <0.01 |
| | 5-cascaded Voxelmorph vs. joint model | 0.70 | 0.97 | 0.44 |
| Right submandibular gland | Elastix vs. joint model | 0.13 | 0.46 | 0.15 |
| | Voxelmorph vs. joint model | <0.01 | 0.04 | <0.01 |
| | 5-cascaded Voxelmorph vs. joint model | 0.97 | 0.92 | <0.01 |
| Spinal cord | Elastix vs. joint model | 0.06 | 0.46 | 0.62 |
| | Voxelmorph vs. joint model | <0.01 | 0.08 | <0.01 |
| | 5-cascaded Voxelmorph vs. joint model | 0.74 | 0.91 | 0.95 |

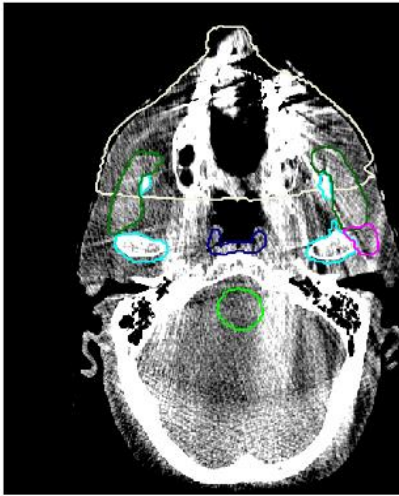 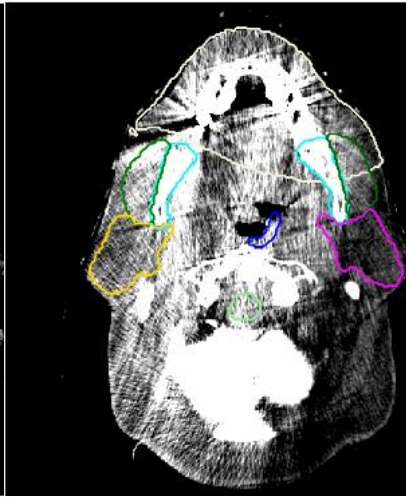 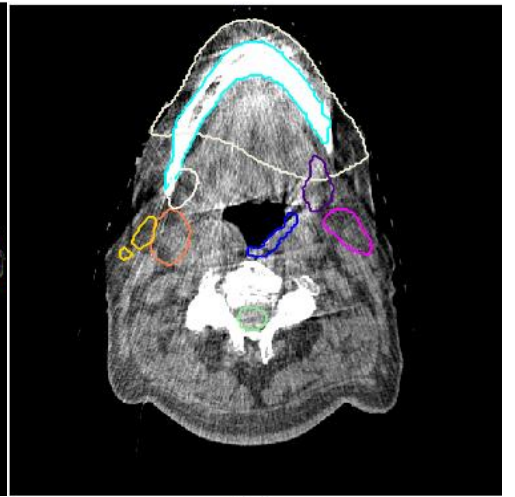

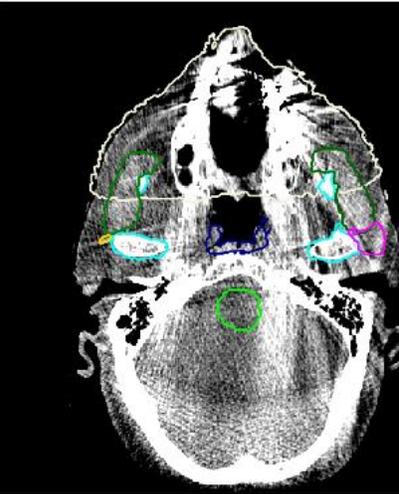 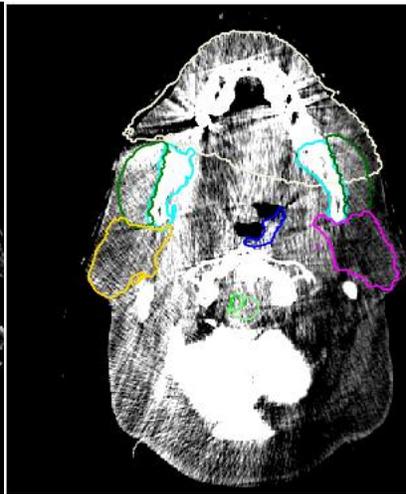 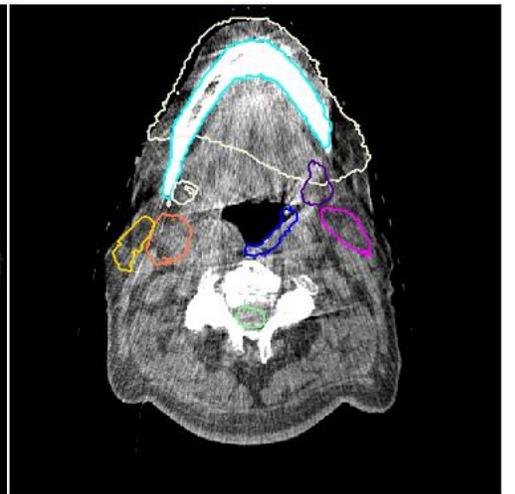

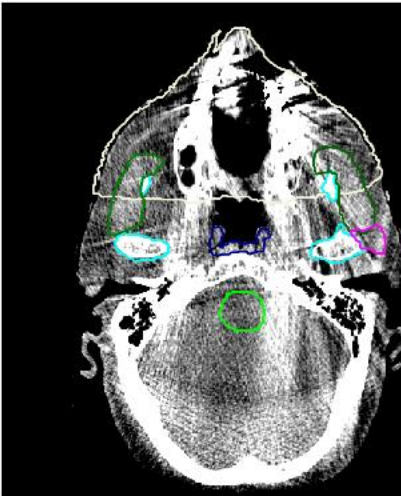 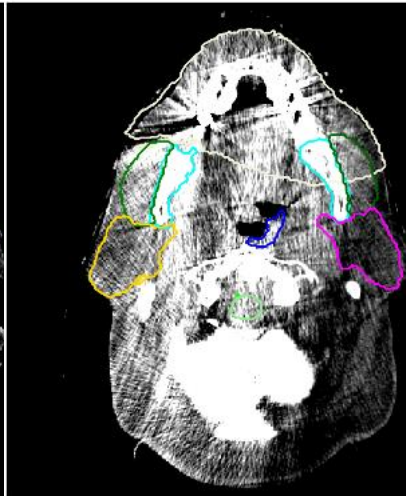 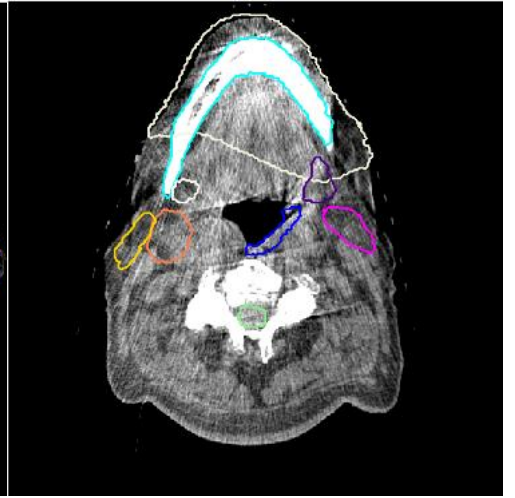

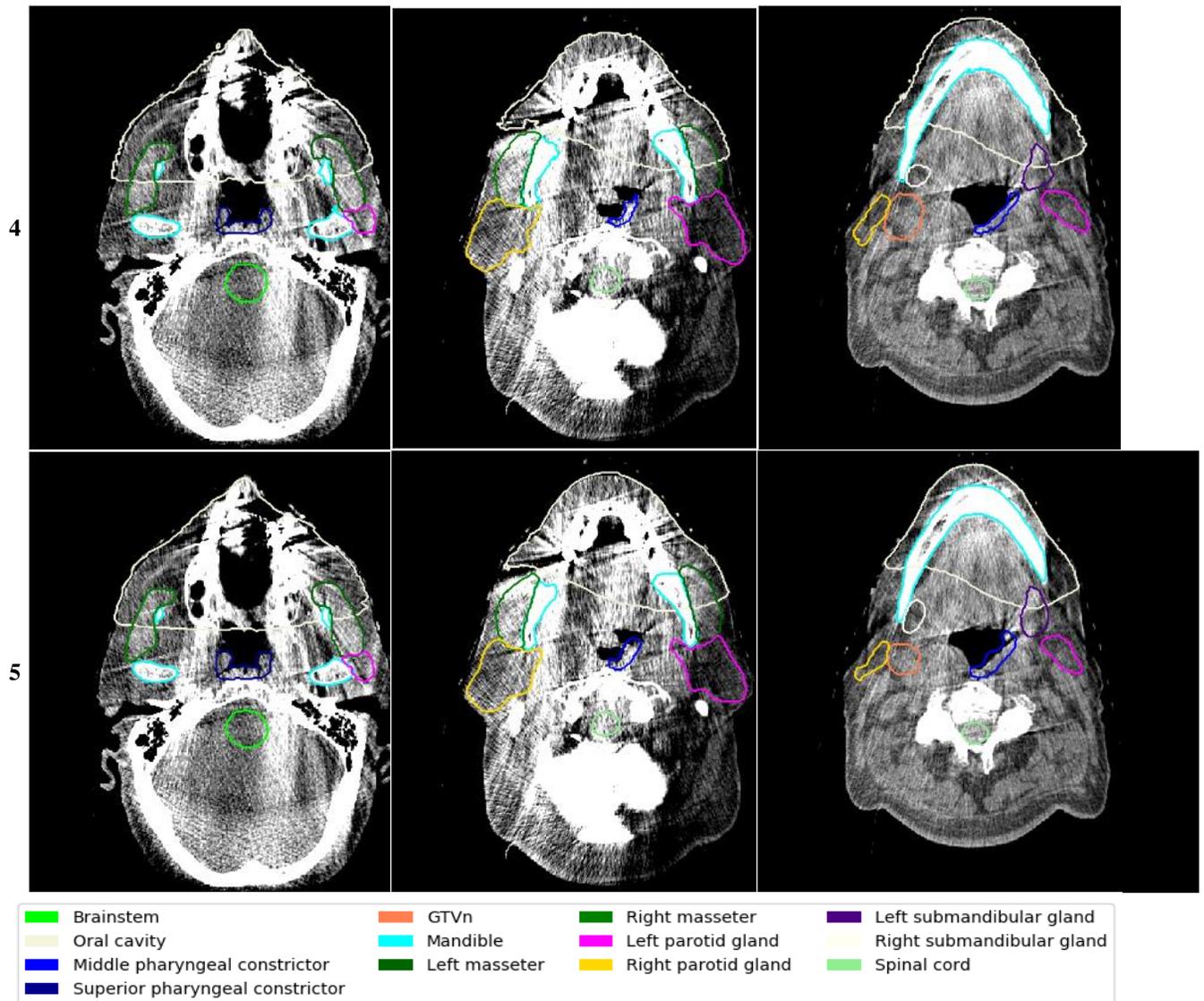

**Figure 4.** The segmentations on CBCT by 1: Elastix, 2: Voxelmorph, 3: 5-cascaded Voxelmorph, 4: joint model, and 5: ground truth for a test patient on axial view.

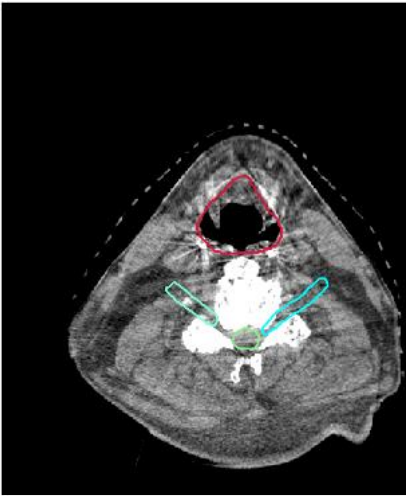 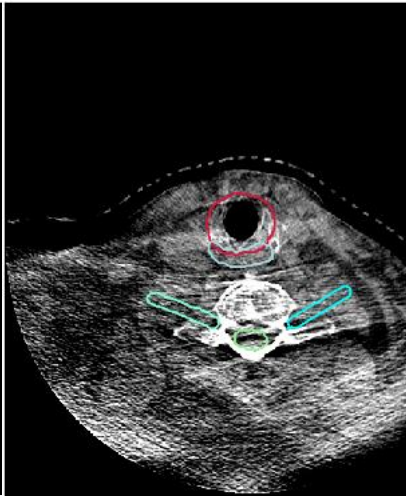 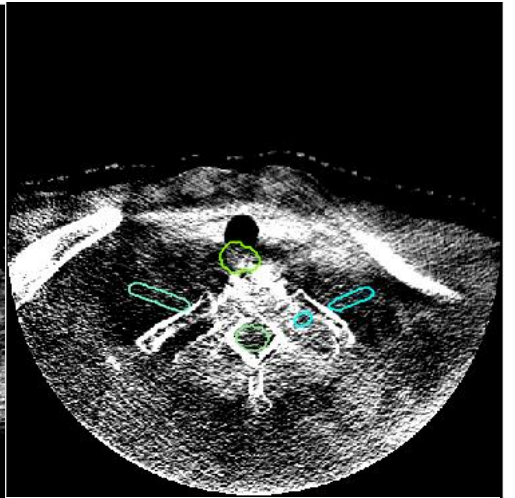
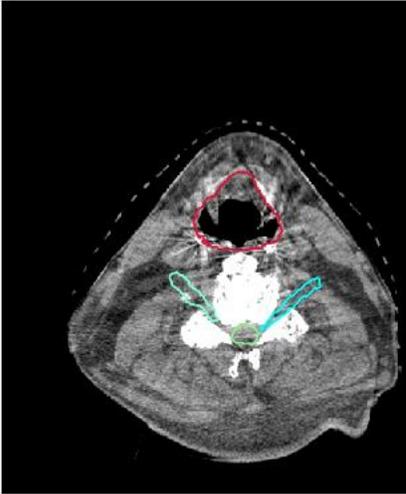 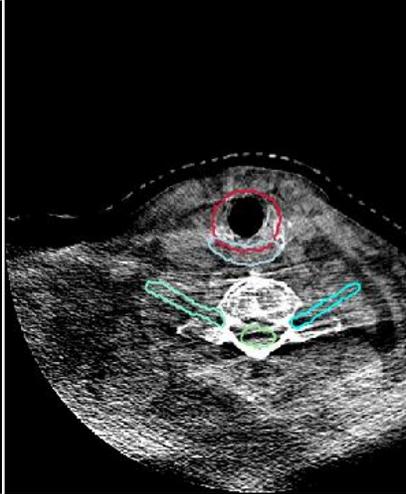 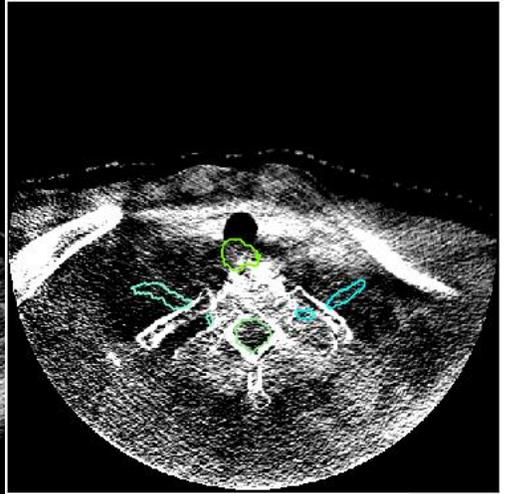
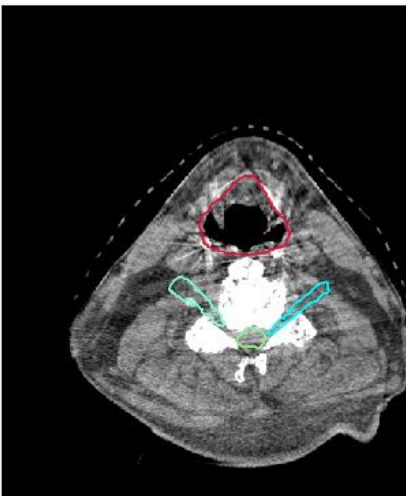 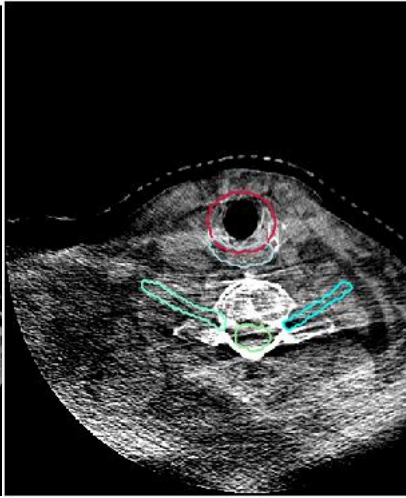 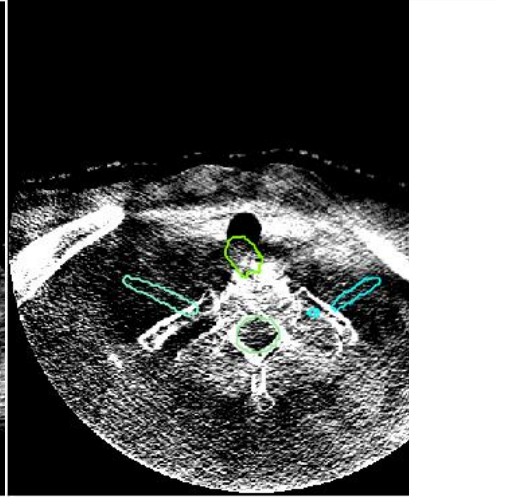

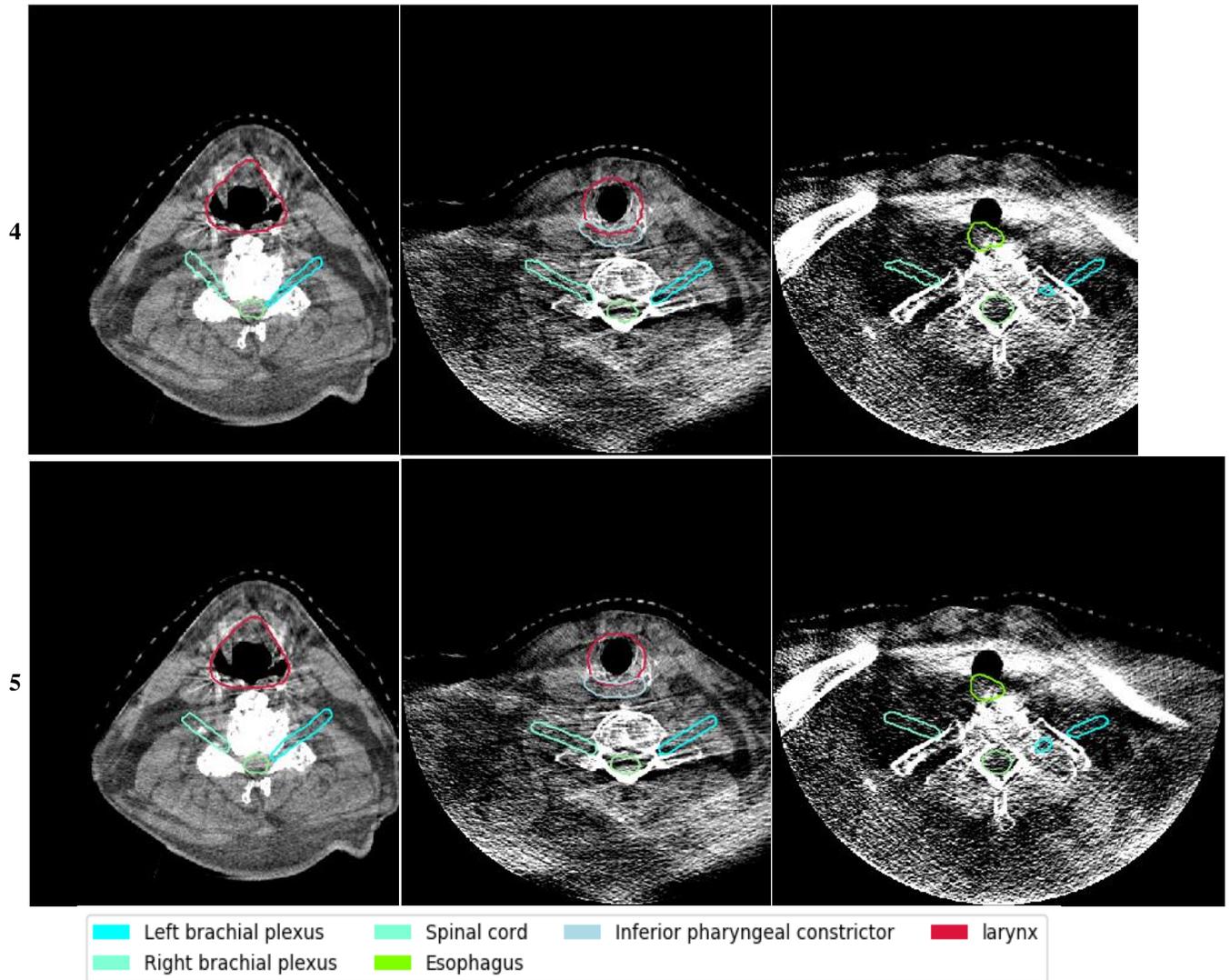

**Figure 5.** The segmentations on CBCT by 1: Elastix, 2: Voxelmorph, 3: 5-cascaded Voxelmorph, 4: joint model, and 5: ground truth for a test patient on axial view.

## 5. Discussion and Conclusion

The sCT images generated by CycleGAN model trained alone and CycleGAN model trained jointly with 5-cascaded Voxelmorph are shown in Figure 6. It shows that sCT generated by the joint model is smoother than sCT generated by CycleGAN alone. This phenomenon confirms with the assumption that adding dpCT in CycleGAN loss function could introduce patient specific knowledge to guide training, while training CycleGAN alone is lack of this information because of unpaired training scheme. Consequently, a better sCT image quality can be achieved by jointly training and thus results in more accurate image registration. Therefore, joint model can outperform 5-cascaded Voxelmorph on some structures.

However, we didn't see the learning based methods surpass the traditional DIR method. For most the structures, our proposed method is comparable to Elastix. With better performance on mandible, esophagus, and superior pharyngeal constrictor, the joint model nevertheless has worse performance on left and right brachial plexus. This is due to sCT over smooth and also structure itself vague on CT images. However, the running time for the joint model after the training can be completed in a minute for each patient, which is much faster than Elastix. In online ART workflow, where time is limited, the deep-learning-based method is more suitable than Elastix.

One limitation of our method is the generalizability problem. According to our previous study, CycleGAN trained on CBCT from one distribution may not work on CBCT from another distribution [15]. Thus, the proposed model needs to be retrained or fine-tuned before its deployment in other institutions.

In conclusion, we developed a learning based DIR method for contour propagation that can be used in ART. The proposed method could generate sCT with correct CT numbers for dose calculation and propagate the contours from pCT to CBCT for treatment re-planning rapidly at the same time, which is a promising tool for external beam online ART.

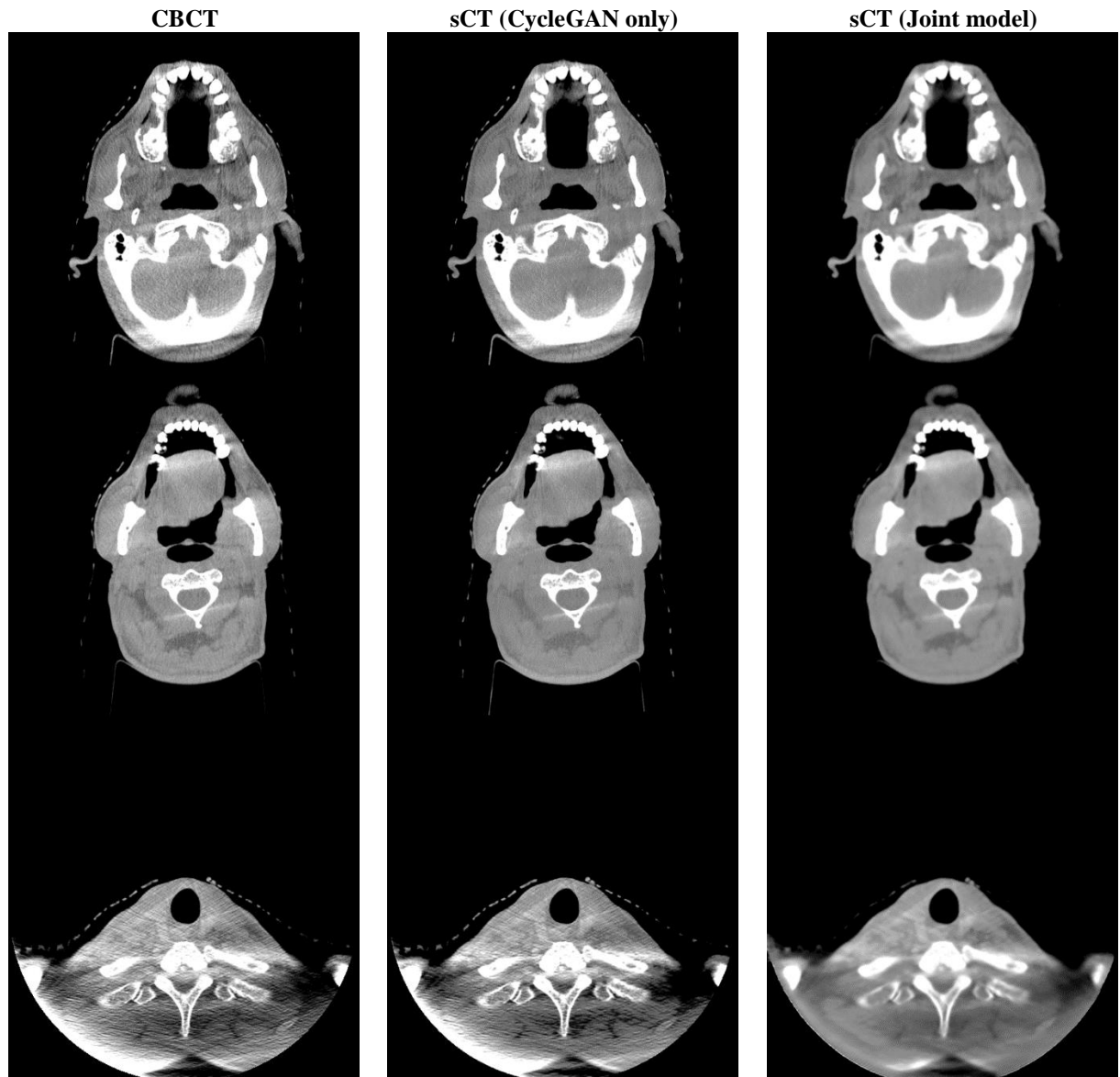

**Figure 6.** Axial view of CBCT and sCT images. From left to right, the images are CBCT, sCT generated by CycleGAN only, and sCT generated by joint model. HU window is (-500,500).

## Data availability

All datasets were collected from one institution and are non-public. According to HIPAA policy, access to the dataset will be granted on a case by case basis upon submission of a request to the corresponding authors and the institution.

## Conflict of interests

The authors declare no competing financial interest. The authors confirm that all funding sources supporting the work and all institutions or people who contributed to the work, but who do not meet the criteria for authorship, are acknowledged. The authors also confirm that all commercial affiliations, stock ownership, equity interests or patent licensing arrangements that could be considered to pose a financial conflict of interest in connection with the work have been disclosed.

## Acknowledgement

We would like to thank the Varian Medical Systems Inc. for supporting this study and Dr. Jonathan Feinberg for editing the manuscript.